\documentclass[twocolumn,english,aps,prl,superscriptaddress,preprintnumbers]{revtex4}
\usepackage[T1]{fontenc}
\usepackage[latin9]{inputenc}
\usepackage{amsmath}
\usepackage{amssymb}
\usepackage{graphicx}

\usepackage{hyperref}

\hypersetup{
    colorlinks=true,
    linkcolor=black,
    citecolor=black,
    filecolor=black,
    urlcolor=black,
}

\makeatletter

\usepackage{amstext}

\usepackage{babel}

\makeatother

\begin{document}

\title{Nonlocal effects: relevance for the spontaneous emission rates of quantum emitters coupled to plasmonic structures}

\author{Robert Filter}
\homepage{http://robertfilter.net}
\email{robert.filter@uni-jena.de}
\affiliation{
Institute of Condensed Matter Theory and Solid State Optics,
Abbe Center of Photonics, Friedrich-Schiller-Universit{\"a}t Jena, D-07743
Jena, Germany}
\author{Christoph B{\"o}sel}
\affiliation{
Institute of Condensed Matter Theory and Solid State Optics,
Abbe Center of Photonics, Friedrich-Schiller-Universit{\"a}t Jena, D-07743
Jena, Germany}
\author{Giuseppe Toscano}
\affiliation{
Institut f{\"u}r Theoretische Festk{\"o}rperphysik, Karlsruhe Institute of Technology, D-76128 Karlsruhe, Germany}
\author{Falk Lederer}
\affiliation{
Institute of Condensed Matter Theory and Solid State Optics,
Abbe Center of Photonics, Friedrich-Schiller-Universit{\"a}t Jena, D-07743
Jena, Germany}
\author{Carsten Rockstuhl}
\affiliation{
Institute of Condensed Matter Theory and Solid State Optics,
Abbe Center of Photonics, Friedrich-Schiller-Universit{\"a}t Jena, D-07743
Jena, Germany}
\affiliation{
Institut f{\"u}r Theoretische Festk{\"o}rperphysik, Karlsruhe Institute of Technology, D-76128 Karlsruhe, Germany}
\affiliation{Institut f{\"u}r Nanotechnologie, Karlsruhe Institute of Technology, 76021 Karlsruhe, Germany}

\begin{abstract}
The spontaneous emission rate of dipole emitters close to plasmonic dimers are theoretically studied within a nonlocal hydrodynamic model.
A nonlocal model has to be used since quantum emitters in the immediate environment of a metallic nanoparticle probe its electronic structure.
Compared to local calculations, the emission rate is significantly reduced. The influence is mostly pronounced if the emitter is located close to sharp edges. We suggest to use quantum emitters to test nonlocal effects in experimentally feasible configurations.
\\ \url{http://dx.doi.org/10.1364/OL.39.006118}
\end{abstract}

\maketitle

Advances in nano-technology permit the investigation of ever smaller gap- and particle sizes of
plasmonic systems \cite{busson2012accelerated,alaee2013deep}. Reaching a truly nanometric regime has given rise to new questions
on how light-matter-interactions have to be properly described.
One of the important insights that could change our understanding of
ultra-small metallic systems is their inherent nonlocal response to electromagnetic excitations \cite{mcmahon2009nonlocal}. The consideration of such nonlocal effects is inevitable whenever the particles are small and small particles are brought in close proximity to each other, e.g. at distances smaller than a few nanometers. In such situations, a series of experimental studies
have shown that local theories fail to properly predict the plasmonic resonances
compared to measurements.
In this case a local theory predicts an excessive redshift of the plasmon resonance that was not found in experiments \cite{liebsch1993surface,tiggesbaumker1993blue,Ciraci2012probing,scholl2013observation}.

Several attempts have been made to properly describe the experimental observations including time-dependent density functional theory (TDDFT) and
the nonlocal hydrodynamic model
(NLHD) \cite{fuchs1987multipolar,rojas1988nonlocal,Llorenc1997core,garcia2008nonlocal,mcmahon2009nonlocal,mcmahon2010optical,
david2011spatial,toscano2012modified,teperik2013quantum,teperik2013robust,toscano2012surface}.
The NLHD is a semiclassical model
based on the Thomas-Fermi theory of the free-electron gas \cite{raza2011unusual}, that has proven to give an accurate description
of the optical response of noble metal nanoparticles \cite{Ciraci2012probing}.
It allows to study plasmonic systems of
nanoparticles of large dimensions that cannot be addressed by means of ab-initio calculations.

On the base of the experimental results obtained from coupled nanoparticles it can be anticipated that such nonlocal effects are not just important if different nanoparticles are brought in close proximity, but also if other nanoscopic systems such as quantum emitters are brought close to metallic nanostructures. This touches plasmonics at its heart, since enhancing light-matter-interactions is a key driver for the further development of subwavelength plasmonic systems \cite{giannini2010controlling,tame2013quantum}.
In this context, it has been already shown how nonlocal effects dramatically lower local electric field enhancements \cite{toscano2012surface}.

\scalebox{.015}{\url{http://robertfilter.net}}\hspace{-0.8mm}In this Letter, we numerically investigate the spontaneous emission
rate of quantum emitters in the vicinity of three-dimensional silver dimers whose nonlocal response is described by the NLHD.
We find that the Purcell factor $F$, i.e. the ratio of the spontaneous emission rate to free space emission, is generally lower as predicted by a local theory.

Within the hydrodynamic model, the nonlocal response of a metal in frequency space is described by the equations \cite{toscano2013nonlocal}
\begin{eqnarray}
\nabla\times\nabla\times\mathbf{E}\left(\mathbf{r}\right)-
\frac{\omega^{2}}{c^{2}}\mathbf{E}\left(\mathbf{r}\right)&=&
\mathrm{i}\omega\mu_{0}\mathbf{j}\left(\mathbf{r}\right)\ \mathrm{,}\\
\frac{\beta^{2}}{\omega\left(\omega+\mathrm{i}\gamma\right)}
\nabla\left[\nabla\cdot\mathbf{j}\left(\mathbf{r}\right)\right]+
\mathbf{j}\left(\mathbf{r}\right)&=&\sigma_\mathrm{D}\left(\mathbf{r}
\right)\mathbf{E}\left(\mathbf{r}\right)\ . \label{eq:math-model}
\end{eqnarray}
The first term in Eq.~\eqref{eq:math-model} is responsible for the nonlocal response.
In the Thomas-Fermi model \cite{halevi1995hydrodynamic}, the nonlocal-response parameter is given by $\beta=\sqrt{3/5}\,v_\mathrm{F}$ with the Fermi velocity $v_\mathrm{F}$. This description is valid for a bulk Drude-type metal with plasma frequency $\omega_p$, damping rate $\gamma$ and corresponding Drude conductivity $\sigma_\mathrm{D}$. It is derived as a first-order-approximation to the hydrodynamic model for the charge density inside a metal \cite{raza2011unusual,toscano2013nonlocal}.
To account for the dispersion of the metal in a simplified manner, the physical parameters in Eq.~\eqref{eq:math-model} are determined at each
frequency of the illumination source. We use the approximations $\gamma\left(\omega\right)\approx\Im\left[\varepsilon\left(\omega\right)\right]\omega^{3}/\omega_{p}^{2}$
and $\omega_{p}\left(\omega\right)\approx\sqrt{1-\Re\left[\varepsilon\left(\omega\right)\right]}\,\omega$, which are valid for
$\gamma\ll\omega$ \cite{johnson1972optical}. For our computations, the local material parameters for silver are
taken from Ref.~\onlinecite{palik1998handbook}.
We further assume that the investigated systems are embedded in fused silica with $\varepsilon_d=2.2$ to hold them in place and to prevent silver oxidation.

To understand why nonlocal effects matter on short length scales, $v_\mathrm{F}$ may
be related to two length scales: the so-called Fermi screening length $\delta_\mathrm{F}=v_\mathrm{F}/\omega_p$ and the Fermi wavelength $\lambda_\mathrm{F}=v_\mathrm{F}/\omega$.
$\delta_\mathrm{F}$ is the characteristic length scale to which the plasmonic surface charge is smeared out.
Throughout this Letter we shall use $v_\mathrm{F}=1.39\cdot10^6\,$m/s \cite{toscano2013nonlocal},
which gives $\delta_\mathrm{F}=1.0\,${\AA} for, say, $\hbar\omega_p = 8.9\,$eV.
$\lambda_\mathrm{F}$ on the other hand defines the wavelength of nonlocal oscillations at a certain excitation frequency $\omega$ \cite{raza2011unusual}. For $\hbar\omega=2.0\,$eV ($\lambda=620\,$nm), $\lambda_\mathrm{F}\approx4.6\,${\AA} such that we may define $\tilde\lambda_\mathrm{F}\equiv 5\,${\AA} as a characteristic nonlocal wavelength.
Both characteristic lengths are on the {\AA}-scale.
So the maximum size of plasmonic particles we are able to investigate within our COMSOL simulations is limited by the requirement of a sub-nm mesh at least close to their surfaces.

We shall analyze the Purcell factor $F$ for a metallic dimer as it usually exhibits a much better radiation efficiency compared to single particles \cite{rogobete2007design}.
We will also use a minimum gap size of $d=6\,$nm which is sufficiently large to neglect electron spill-out effects \cite{Kern2012atomic,teperik2013robust,david2014surface,gtoscanoSC} and to analyze the system in the weak coupling regime \cite{slowik2013strong}, in which $F$ can be calculated within classical electrodynamics as the ratio of the radiated powers with and without nanoparticles, $F=P_\mathrm{rad}/P_\mathrm{rad}^\mathrm{fs}$ \cite{bharadwaj2009optical}.

\begin{figure}[tb]
\centerline{\includegraphics[width=8cm]{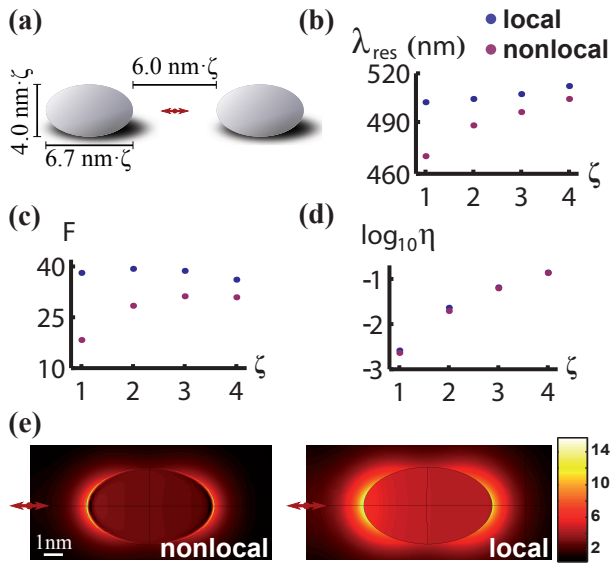}}
\caption{\label{fig:conformal}
(a) Schematic. A dipole is centered in-between two silver spheroids with the given dimensions with conformal factor $\zeta=\{1,2,3,4\}$. The dipole is oriented parallel to the symmetry axis (red double-arrow).
(b) Resonance wavelength $\lambda_\mathrm{res}$ of the dimers for local (blue) and nonlocal (magenta) calculations for a maximal Purcell factor $F(\lambda)$.
(c) Purcell factor $F(\lambda_\mathrm{res})$ at resonance. For $\zeta=1$, the local model predicts more than twice the spontaneous emission rate enhancement than in the NLHD. For $\zeta=4$, the difference is roughly only $20\,$\%.
(d) Radiation efficiency $\eta$ for the dipole within the lossy silver spheroid dimer.
(e) Absolute value of the scattered electric field inside one metallic particle for nonlocal and local calculations.
}
\end{figure}
The first interesting question is how $F$ varies with dimer size.
We would expect that nonlocal effects provide only a minor contribution for sufficiently big particles.
In Fig.~\ref{fig:conformal}, a conformal factor $\zeta$ is used to increase any scale of a dimer made of silver spheroids,
see Fig.~\ref{fig:conformal} (a) for a schematic.
The conformal factor was limited to $\zeta\leq 4$ because of the large computational demand of the calculations ($\zeta=4$ implies a dimer size of almost $80\,$nm).
As shown in Fig.~\ref{fig:conformal} (b), a blueshift of more than $30\,$nm between nonlocal and local descriptions can be observed for $\zeta=1$, which reduces to less than $10\,$nm for $\zeta=4$.
The larger blueshift for lower $\zeta$ can be explained by a large relative increase in the effective distance between the surface charges for smaller separations which leads to a weaker hybridization of the particles \cite{toscano2012modified}.
The found shifts are comparable to those identified for spherical gold dimers under plane-wave excitation \cite{david2011spatial}.
Most importantly, we also find a strong effect of the respective models on the Purcell factor $F$ (Fig. \ref{fig:conformal} (c)).

One may pose the question if the difference in $F$ may be explained by a difference in radiation efficiency $\eta$.
In the weak coupling regime, $\eta$ can be calculated
as the ratio of the radiated power $P_\mathrm{rad}$ compared to the total emitted power that is given by
a sum of $P_\mathrm{rad}$ and the power that is dissipated inside the dimer, $P_\mathrm{diss}$: $\eta=P_\mathrm{rad}/(P_\mathrm{rad}+P_\mathrm{diss})$ \cite{bharadwaj2009optical}.
Fortunately, the dissipated power inside of the dimer can be calculated as in the case of a local dispersive medium \cite{landau1960electrodynamics,toscano2013nonlocal}.
Due to different particle sizes, $\eta$ changes by orders of magnitude for varying $\zeta$, as illustrated in Fig.~\ref{fig:conformal} (d).
Further, the nonlocal efficiencies $\eta$ are always less than the local ones but their difference is significantly lower than the differences in $F$. Hence, the difference in $F$ cannot be explained by a difference in $\eta$. This important result also holds for the other investigated geometries.

The reason for the different Purcell factors lies in the distribution of the electric field along the spheroids.
Figure \ref{fig:conformal}~(e) shows that in the NLHD model the field is strongly enhanced in a very tiny region inside the spheroid close to its surface which we shall term screening region.
On the contrary, an almost homogeneous field inside the particles is found for local calculations and the field is strongest outside the dimer.
The confinement in the NLHD model leads to a decrease of the electric field in the dimer gap and thus a weaker interaction with the dipole emitter and hence a lower Purcell factor. Especially close to the surface the influence of nonlocal effects is pronounced such that we can expect drastic changes of $F$ there \cite{larkin2004dipolar,johansson2005surface}.

\begin{figure}
\centerline{\includegraphics[width=8cm]{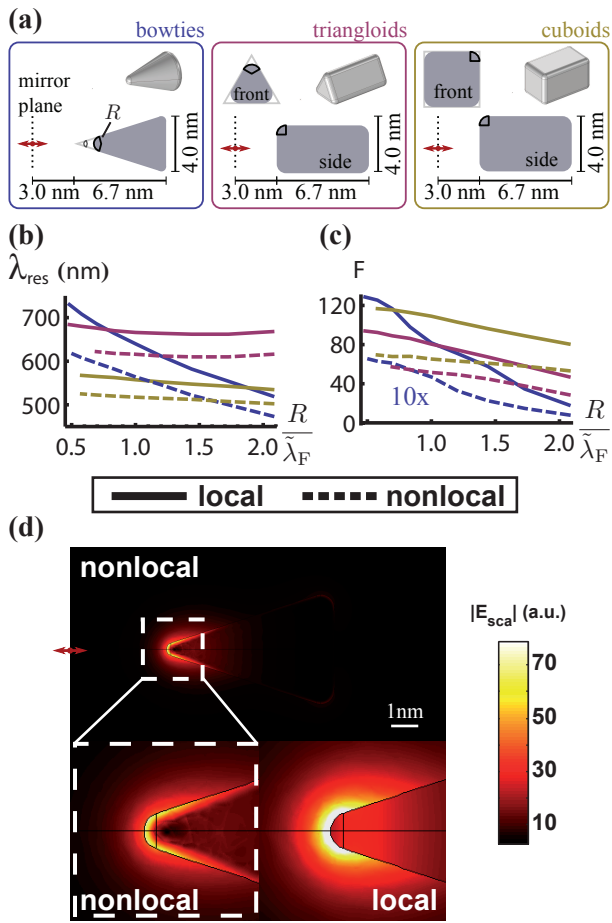}}
\caption{\label{fig:edges}
Investigation of Purcell factors with respect to the local radius of curvature for differently shaped dimers.
(a) Investigated geometries: bowties (blue box), triangloids (magenta box) and cuboids (yellow box) - only one of the symmetric dimer elements is shown (colors apply for (b) and (c) too).
The dipole is centered in-between the nanoparticles with a total gap of $6\,$nm.
The bowties are spherically terminated with varying radius $R$, whereas for the triangloids and cuboids any edge is rounded with a radius $R$.
(b) Resonance wavelength $\lambda_\mathrm{res}$ of the various dimers obtained from local (solid lines) and NLHD (dashed lines) calculations with respect to a maximum Purcell factor $F(\lambda)$ depending on $R/\tilde{\lambda}_\mathrm{F}$.
(c) Purcell factor calculations. The bowtie results are scaled by a factor of $10$ for a better visibility.
(d) The norm of the scattered electric field (a.u.) for a side view cut through a centered layer for a bowtie element with
$R/\tilde{\lambda}_\mathrm{F}=\left(5/6\right)^4\approx0.48$. The values of the electric field in the local plot have been cropped
such that any value $|\mathbf{E}_\mathrm{sca}|>75.5$ appears in white color.
}
\end{figure}

With the latter investigations of conformally rescaled spheroid dimers we found indications that the Purcell factor generally decreases because of nonlocal effects. Naturally, the local radius of curvature of the investigated spheroids were much bigger than any characteristic nonlocal length scale.

We may now generalize our findings for a class of dimers with different shapes for which we vary the local radius of curvature $R$ around the characteristic $\tilde\lambda_\mathrm{F}=5${\AA}.
The geometries of the dimers we shall analyze are depicted in Fig.~\ref{fig:edges} (a).
Most notably, bowties on the one side and triangloids and cuboids on the other side differ by their respective rounding schemes: Whereas in the case of bowties their termination approaches the dipole with decreasing $R$, the distance to the front surfaces of triangloids and cuboids always remains $3\,$nm ($d=6\,$nm). The bowties exhibit varyingly sharp edges near the emitting dipole whereas the triangloids and cuboids exhibit edges mainly on their sides.
These different rounding
schemes permit the investigation of NLHD effects for the near-field coupling and radiation properties separately:
In the case of a sharp edge facing the emitter in the near-field (bowties),
their mutual coupling is strongly altered.
If the sharp edges are placed on the sides of the particles (triangloids and cuboids), a change of their local
curvature does not drastically alter the mutual coupling but influences the particle's radiation properties only.

The resonance wavelengths $\lambda_\mathrm{res}$ of the respective dimers
exhibit a blueshift in the NLHD calculations compared to the local ones
(Fig.~\ref{fig:edges} (b)).
The bowties exhibit a blueshift of more than $100\,$nm for the smallest $R$, which reduces to about $50\,$nm for $R/\tilde{\lambda}_\mathrm{F}> \left(5/6\right)^{-3}\approx1.7$. For the triangloids and cuboids, the blueshift is not as pronounced with a maximum of around $50\,$nm and $40\,$nm, respectively. In these cases we find only a moderate (cuboids) or no significant change (triangloids) in the blueshift for varying $R$.
It can also be seen that the change of $R$ hugely influences the resonance frequency of the bowties, whereas the values remain relatively unchanged for triangloids and cuboids.
We observe a strong redshift for decreasing $R$ since the distance between the metal particles is lowered, although the involved metallic volume change is extremely small.

In Fig.~\ref{fig:edges} (c) a general decrease of $F$ for NLHD calculations can be observed in agreement with Fig.~\ref{fig:conformal} (d).
The results for $\lambda_\mathrm{res}$ suggest that $F$ should strongly change for the bowties but should remain constant for the other geometries.
In case of the bowties, $F$ is strongly changing for varying $R$ because the distance to the dipole appreciably affects the coupling between the two subsystems.
For the other geometries, the absolute change in $F$ is much less.
The decrease in $F$ might be simply attributed to a changed particle volume since their mutual distance remains constant.
Noteworthy the ratio of the Purcell factors obtained by the local and nonlocal calculations
is only marginally influenced by the edge radii $R$.
We find this ratio to be about two for small bowties and small spheroids (Fig. \ref{fig:conformal}~(c)), whereas triangloids and cuboids exhibit a factor of about $1.5$. Hence, the influence of nonlocal effects is more pronounced for structures with sharp edges close to the emitter (spheroids and bowties).
From Fig.~\ref{fig:edges}~(d) it is evident how the near-field of a bowtie drastically alters in the nonlocal case.

It can be anticipated that the Purcell factor can be additionally affected by the actual placement of the dipole. We have made computations for a dipole that is displaced from the center of the cuboid dimer
but still lies in the mirror plane. We found that $F$ is influenced by the very placement but that the change for both models are very much the same.
The reason is that the investigated dimers are so small that the emission enhancement can be readily explained by a coupling to the dipolar mode of the dimer.
Then, $F$ only depends on the coupling of the emitting dipole to that mode \cite{filter2014nanoantennas}, which is changed by a varying location of the dipole.
For larger structures, however, there may be more modes that need to be considered. Since these modes might exhibit a different influence of nonlocal effects, suitably placed quantum emitters might be used to probe the general influence of nonlocal effects for extended plasmonic structures.


In conclusion we have investigated how size and shape of metallic dimers affect the spontaneous emission rate of dipoles placed in the gap centre.
In analogy to a lower field enhancement in nanometric gaps \cite{toscano2013nonlocal} we found that nonlocal effects generally reduce the
achievable spontanteous emission rate enhancement $F$ for metallic dimers.
We confirm a general blueshift of the dimer resonances with respect to a close-by emitter that strongly depends on the shape of the dimers.
The spontaneous emission rate is mostly affected by a nonlocal material response if a sharp edge is close to the emitting dipole. The effect is less pronounced if an edge is present at less prominent positions.
Recent studies have shown that configurations with nanometric gap sizes and precise quantum emitter placement are possible \cite{busson2012accelerated}. Hence, quantum emitters may be used as local probes to deepen our knowledge of nonlocal effects as their influence is not a minor disturbance but a main factor for the overall emission.

We hope that this work will be helpful for appropriately designing hybrid quantum and plasmonic systems to pave the way for efficient future applications in quantum communication and nonclassical light sources.


We thank the German Federal Ministry of Education and Research(PhoNa) and the Thuringian State Government(MeMa) for their support.


\end{document}